\documentclass[11pt]{article}
\usepackage{moriond,epsfig}

\def\Journal#1#2#3#4{{#1} {\bf #2}, #3 (#4)}

\def\NPB{{\em Nucl. Phys.} B}
\def\PLB{{\em Phys. Lett.}  B}
\def\PRL{\em Phys. Rev. Lett.}
\def\PRD{{\em Phys. Rev.} D}

\def\be{\begin{equation}}
\def\ee{\end{equation}}
\def\bea{\begin{eqnarray}}
\def\eea{\end{eqnarray}}

\begin{document}
\vspace*{4cm}
\title{Heavy Flavour Production at $\sqrt{s}=7$ TeV}

\author{ E. Aguil\'o }

\address{Physik-Institut, Universit\"at Z\"urich, Winterthurerstr. 190,\\
CH-8057 Z\"urich, Switzerland}

\maketitle\abstracts{
The measurements of b quark, quarkonium and exotic state production performed with the ATLAS and CMS experiments at $\sqrt{s}=7$ TeV are presented. The {b}-quark production cross section is measured both in inclusive and fully reconstructed B hadron decays. The results are compared with QCD expectations at tree-level and NLO.}

\section{Introduction}

The data analyzed for the presented results were collected by the multi-purpose experiments ATLAS~\cite{atlas_2008} and CMS~\cite{cms_2008} at the LHC, which provides since Spring 2010 proton-proton collisions at  $\sqrt{s}$ = 7 TeV. The collected luminosity by each experiment was close to 40 pb$^{-1}$ in 2010 and 5 fb$^{-1}$ in 2011.

The study of heavy quark production cross section in high-energy hadronic interactions plays a critical role as precision tests of next-to-leading order (NLO) Quantum Chromodynamics (QCD) calculations~\cite{hq_1988} at a higher energy scale than before. Measurements of {\it b}-hadron production at the higher energies provided by the Large Hadron Collider (LHC), which are possible thanks to the large $b\bar{b}$ cross section at $\sqrt{s}$ = 7 TeV, represent an important test of theoretical calculations~\cite{had_1998,had_2008}. In addition, a good understanding of {\it b}-quark production is necessary, since it is an important background to several other analyses, {\it i.e.} top quark physics, Higgs or Supersymmetry searches, {\it etc}. These measurements also serve as a validation of the tracking and muon systems.

Both experiments have produced several results on several heavy flavor production subjects, which can be divided in the following three categories: quarkonium production, exclusive heavy flavor hadron cross-section measurements and inclusive b$\bar{\rm b}$ cross-section measurements using b-tagged jets or muons. The latest results on each category will be presented here.

\section{Observation of the $\chi_{\rm b}$(3P)}

\begin{figure}[h]
\begin{center}
\psfig{figure=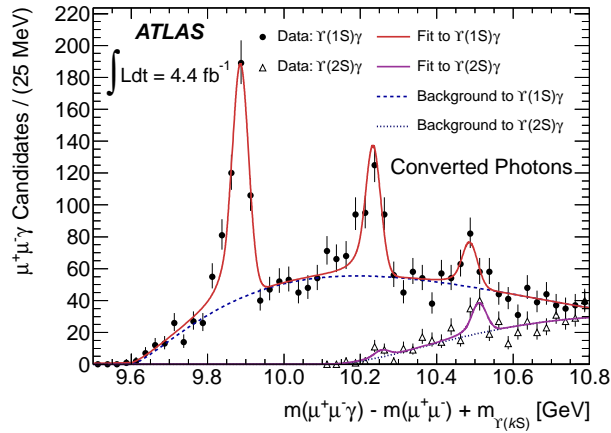,height=6cm}
\end{center}
\caption{The mass distributions of the $\chi_{\rm b} \to \Upsilon{\rm (kS)}\gamma$ (k = 1, 2) candidates formed using photons which have converted and have been reconstructed in the tracker. Data are shown before the correction for the energy loss from the photon conversion electrons due to bremsstrahlung and other processes. The data for decays of $\chi_{\rm b} \to \Upsilon{\rm (1S)}\gamma$ and $\chi_{\rm b} \to \Upsilon{\rm (2S)}\gamma$ are plotted using circles and triangles respectively. Solid lines represent the total fit result for each mass window. The dashed lines represent the background components only.}
\label{fig:chib}
\end{figure}

The $\chi_{\rm b}$ mesons are the P-wave function excitation of b$\bar{\rm b}$ quark system. They decay radiating a photon into the $\Upsilon$(1S) or the $\Upsilon$(2S). These mesons can appear in different spin projections, resulting in a hyperfine splitting of the spectrum. The ATLAS experiment has analyzed the whole 2011 data sample reconstructing $\Upsilon\to\mu^+\mu^-$, and matching them to either calorimeter reconstructed photons or converted photons, from the e$^+$e$^-$ tracks reconstructed in the tracker. Calorimeter photons can be reconstructed more efficiently than converted photons, but in addition to the fact that the converted photons allow the reconstruction of lower energy photons, they also have better energy resolution than those reconstructed in the calorimeter. The spectrum when using converted photons can be seen in Fig.~\ref{fig:chib}, where the right-most peak corresponds to a never observed before $\chi_{\rm b}$ state: the $\chi_{\rm b}$(3P). The spectrum is fitted to crystal ball functions for the signal peaks, including the hyperfine mass splitting structure predicted by Ref.~\cite{chib_theo}, and an empirical function for the background. The $\chi_{\rm b}$(3P) peak significance is larger than 6 standard deviations and the measured mass barycenter is 10.530 $\pm$ 0.005 (stat.) $\pm$ 0.009 (syst.) GeV~\cite{atlas_chib}.

\section{$\Lambda_{\rm b}$ production cross section}

\begin{figure}[h]
\begin{center}
\psfig{figure=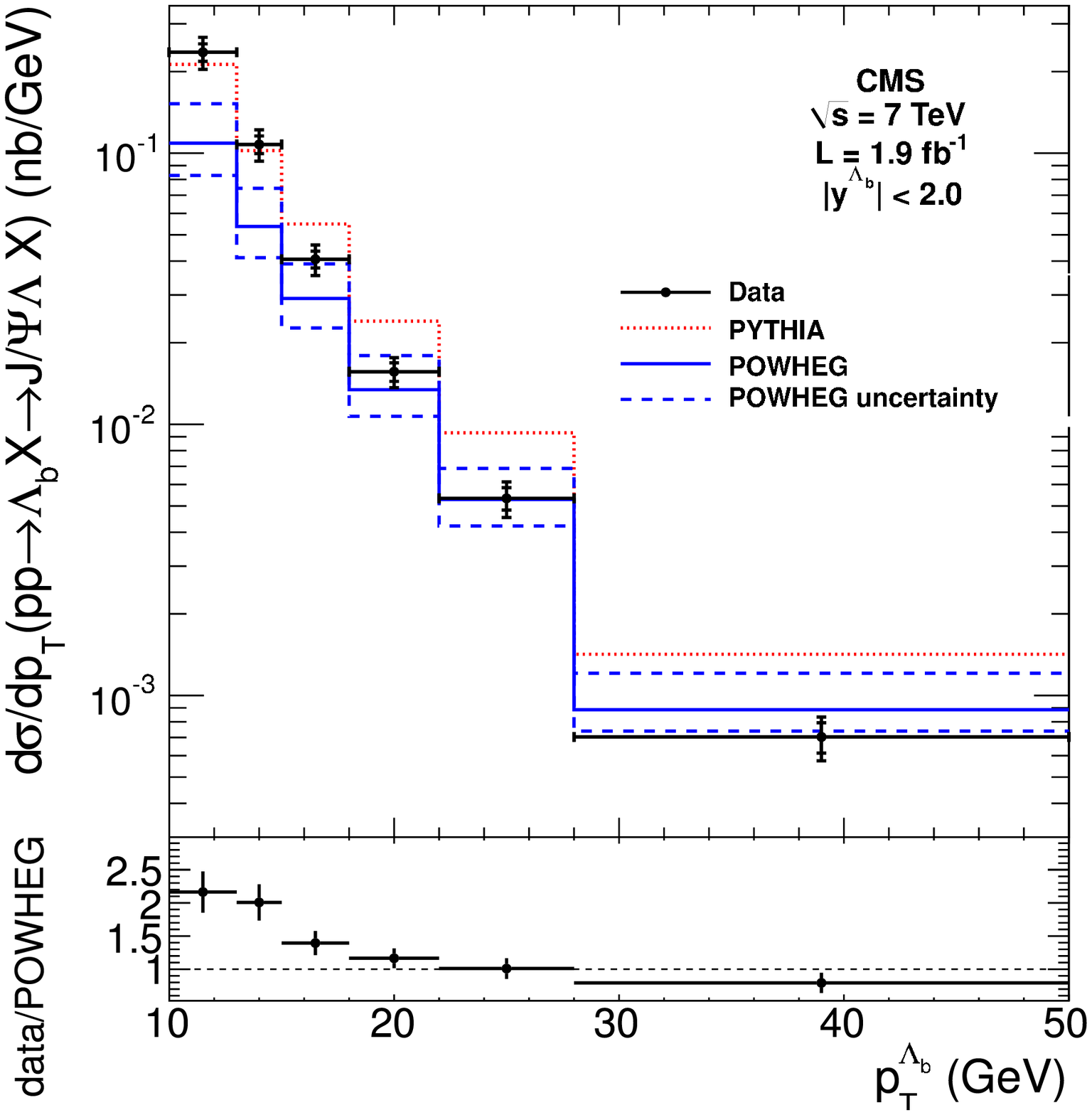,height=7cm}
\psfig{figure=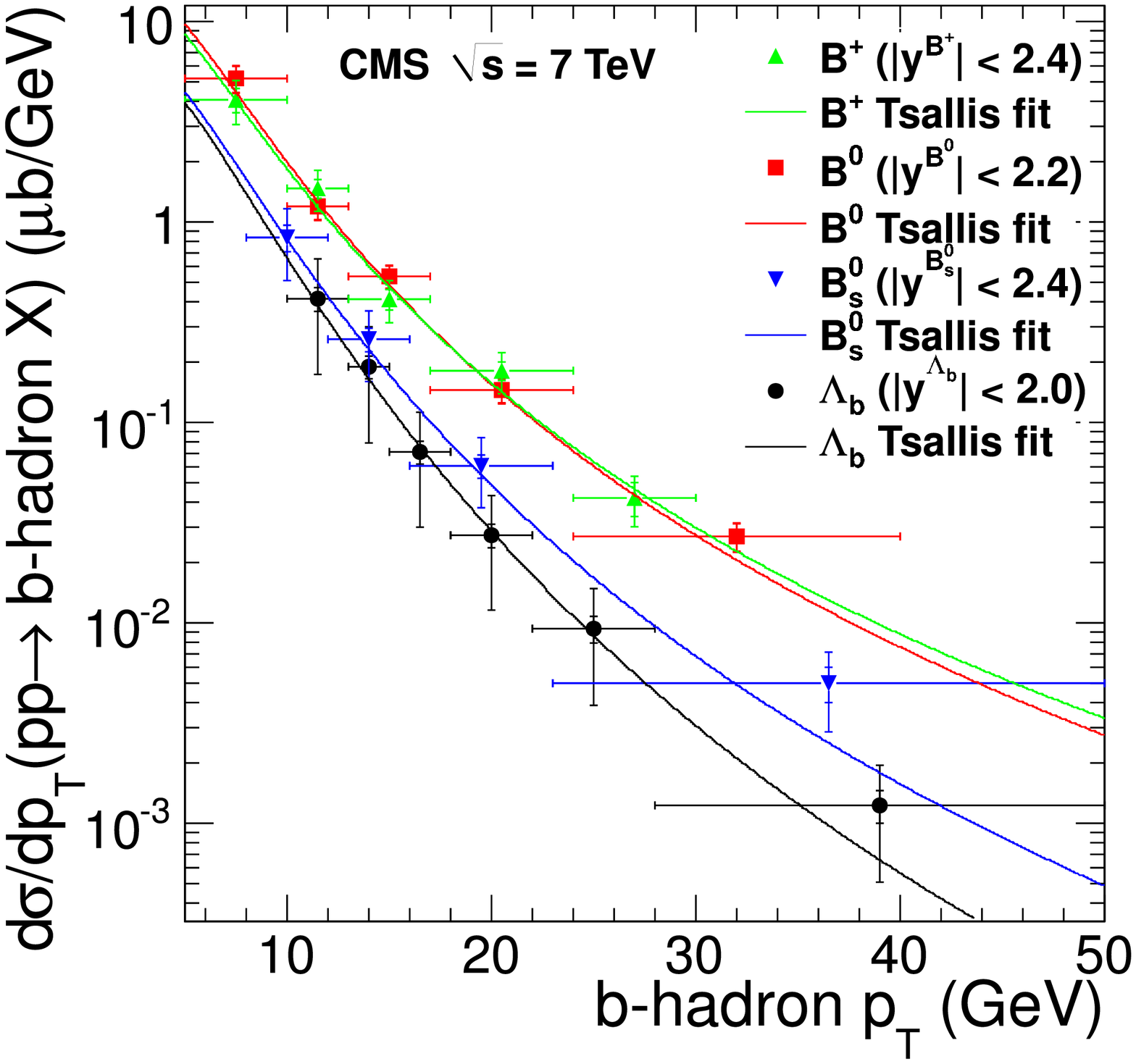,height=7cm}
\caption{(Left) Upper: measured differential cross sections times branching fraction vs. p$_{\rm T}^{\Lambda_{\rm b}}$ compared to the theoretical predictions from PYTHIA and POWHEG. The inner error bars correspond to the statistical uncertainties and the outer ones represent the uncorrelated systematic uncertainties added in quadrature to the statistical uncertainties. The dashed lines show the uncertainties on the POWHEG predictions. Overall uncertainties of 2.2\% for the luminosity and 1.3\% for the J/$\psi\to\mu^+\mu^-$ and $\Lambda\to{\rm p}\pi$ are not shown, nor is the 54\% uncertainty due to $\mathcal{B}(\Lambda_{\rm b}\to {\rm J/}\psi\Lambda)$ for the PYTHIA and POWHEG predictions. Lower: The ratio of the measured values to the POWHEG predictions. The error bars include the statistical and uncorrelated systematic uncertainties on the data and the uncertainties affecting only the distribution shapes on the POWHEG predictions. (Right) Comparison of b-hadron production rates versus hadron p$_{\rm T}$, where the inner error bars correspond to the bin-to-bin uncertainties, while the outer error bars represent the bin-to-bin plus normalization uncertainties added in quadrature. The large normalization uncertainties for $\Lambda_{\rm b}$ and B$_{\rm s}$ are dominated by the poorly measured $\Lambda_{\rm b}\to{\rm J/}\psi\Lambda$ and B$_{\rm s}\to{\rm J/}\psi\phi$ branching fractions for the decay channels used in the analysis.}
\label{fig:lambdaB}
\end{center}
\end{figure}

CMS has measured the $\Lambda_{\rm b}$ baryon production differential cross section~\cite{cms_lambdaB} in transverse momentum (p$_{\rm T}$) and rapidity ($y$) using the decay chain $\Lambda_{\rm b}\to{\rm J/}\psi\Lambda$, J/$\psi\to\mu^+\mu^-$, and $\Lambda\to{\rm p}\pi$ with 1.9 fb$^{-1}$ of the 2011 data sample. As seen in Fig.~\ref{fig:lambdaB} (left), the measured differential cross section shows a steeper slope than the Monte Carlo (MC) predictions. The production ratio between $\Lambda_b$ and $\bar{\Lambda}_b$ is also measured and no significant deviations from theory are found over the measured p$_{\rm T}$ and $y$ ranges. In Fig.~\ref{fig:lambdaB} (right) this result is compared to the differential cross sections of the B$^+$, B$^0$ and B$_{\rm s}$, as measured by CMS.

\section{Production of heavy flavor with b-jets and with muons}

\begin{figure}[h]
\begin{center}
\psfig{figure=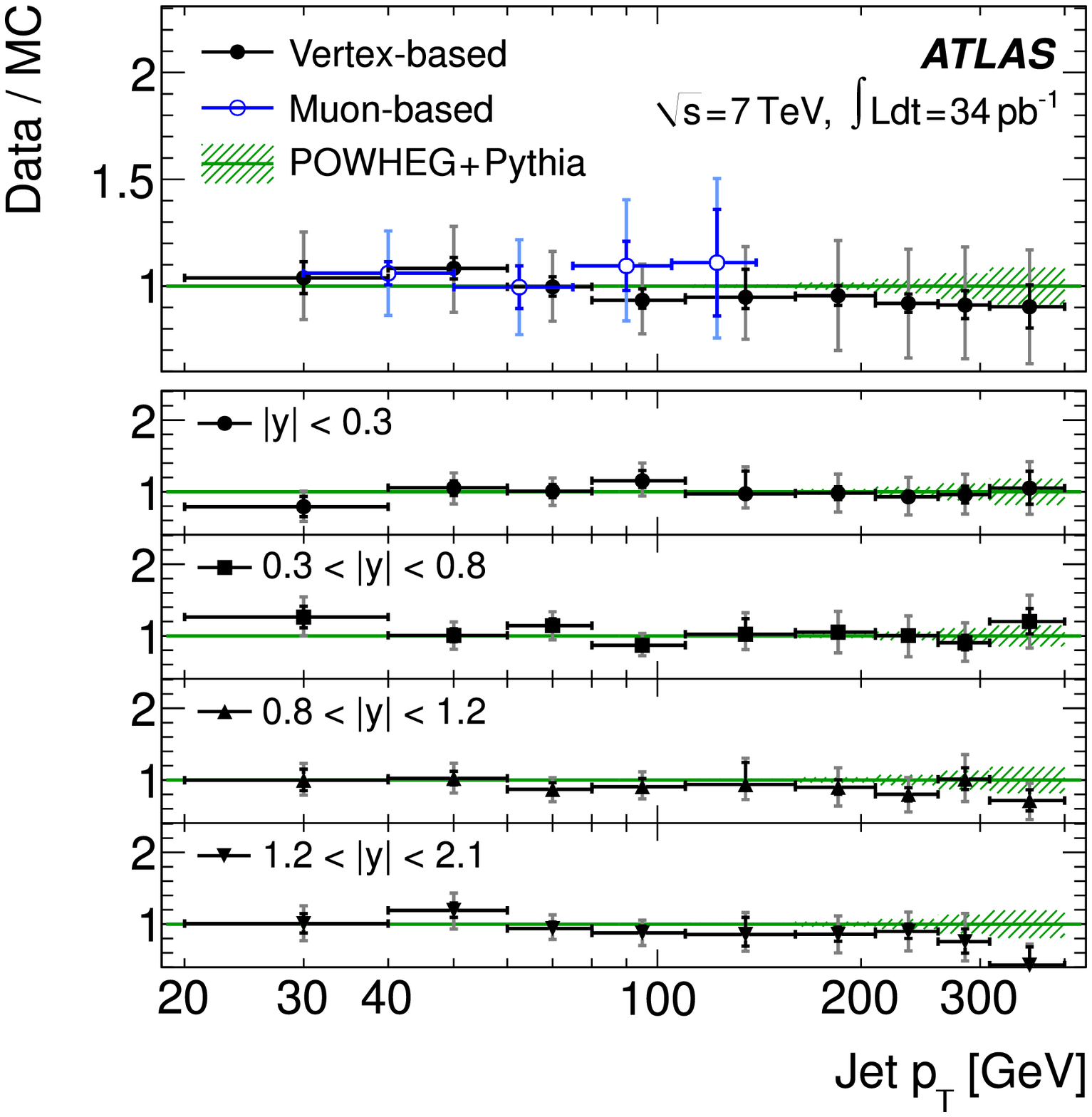,height=7cm}
\psfig{figure=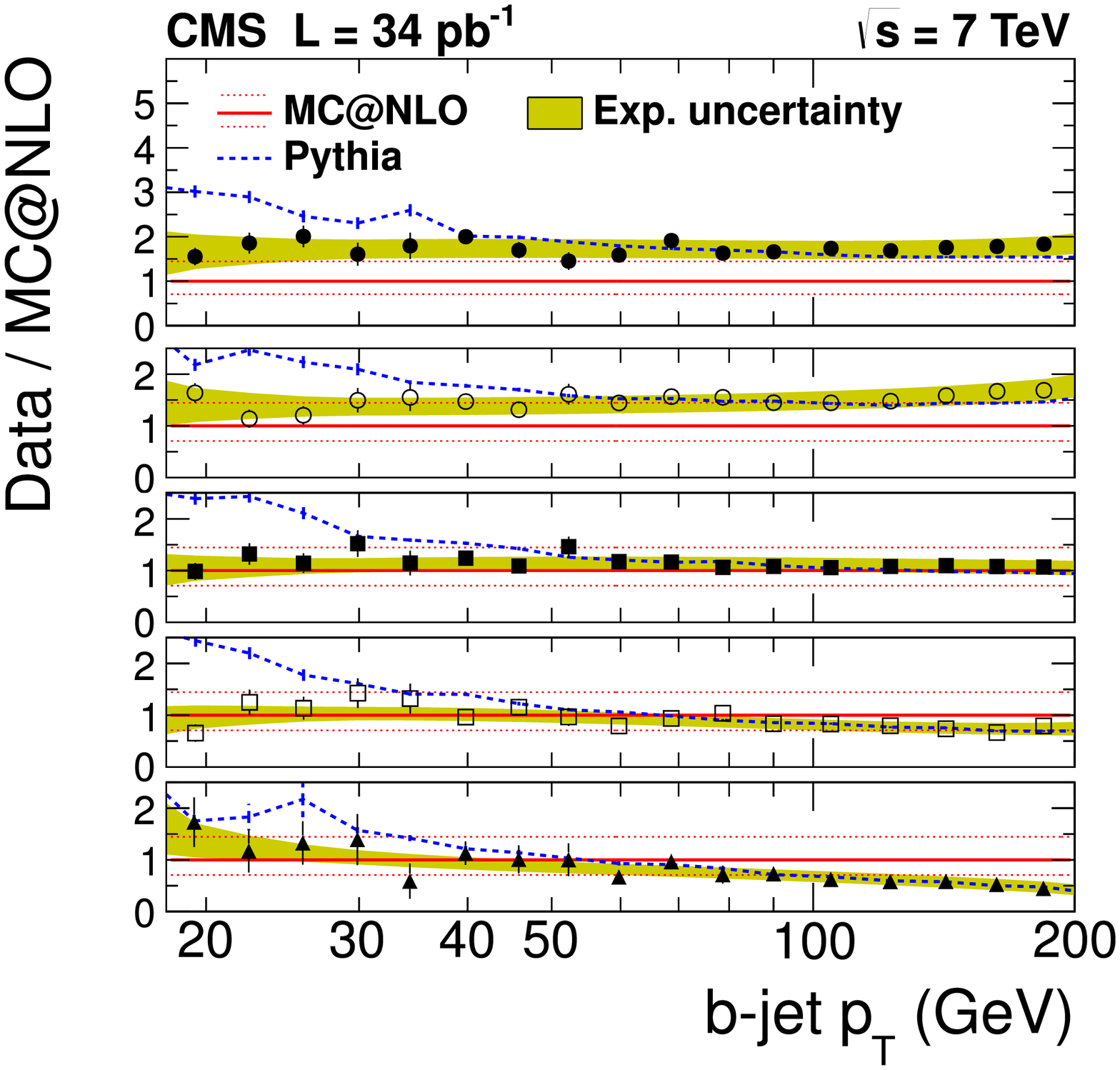,height=7cm}
\caption{(Left) Ratio of the ATLAS measured cross sections to the theory predictions of POWHEG. In the region where the displaced vertex based measurement overlaps with the muon "p$_{\rm T}$ rel" measurement both results are shown. The top plot shows the full $y$ acceptance, while the four smaller plots show the comparison for each of the $y$ ranges separately. The data points show both the statistical uncertainty (dark colour) and the combination of the statistical and systematic uncertainty (light colour). The shaded regions around the theoretical predictions reflect the statistical uncertainty only. (Right) CMS measured b-jet cross section shown as a ratio to the MC@NLO calculation, for the ranges $|y|<0.5$, $0.5<|y|<1$, $1<|y|<1.5$, $1.5<|y|<2$, $2<|y|<2.2$. The experimental systematic uncertainties are shown as a shaded band and the statistical uncertainties as error bars. The MC@NLO uncertainty is shown as dotted lines. The PYTHIA prediction is also shown as a dashed line.}
\label{fig:inc}
\end{center}
\end{figure}

\begin{figure}[h]
\begin{center}
\psfig{figure=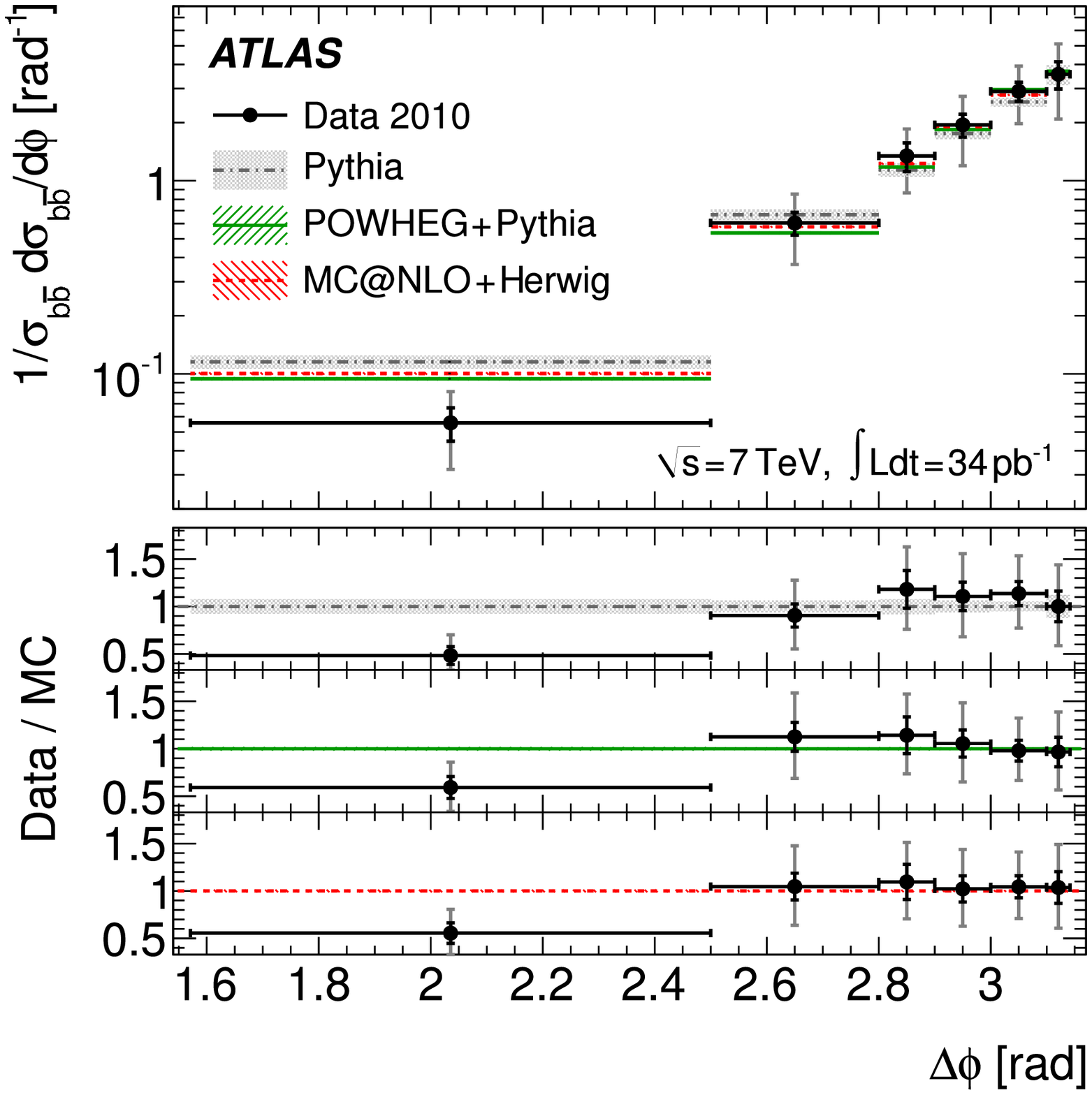,height=7cm}
\psfig{figure=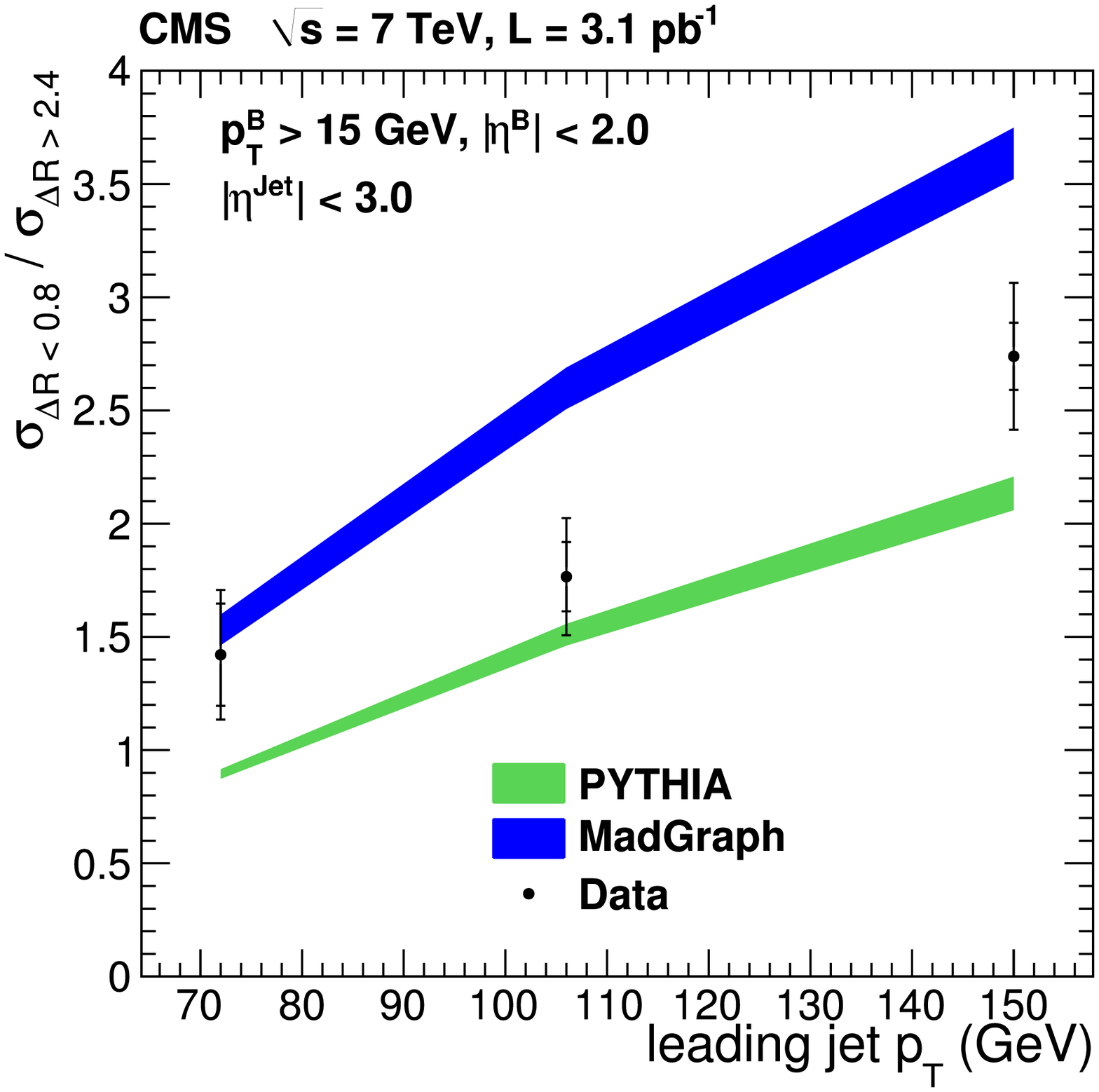,height=7cm,angle=90}
\caption{(Left) The ATLAS b$\bar{\rm b}$-dijet cross section as a function of the azimuthal angle difference between the two jets for b-jets with p$_{\rm T}$ $>$ 40 GeV, $|y|$ $<$ 2.1 and a dijet invariant mass of m$_{\rm jj}$ $>$ 110 GeV. The data are compared to the theory predictions of Pythia, POWHEG and MC@NLO. The shaded regions around the MC predictions reflect the statistical uncertainty only. (Right) ratio between the BB production cross sections in $\Delta$R $<$ 0.8 and $\Delta$R $>$ 2.4 as a function of the leading jet p$_{\rm T}$. For the data points, the error bars show the statistical (inner bars) and the total (outer bars) errors. Also shown are the predictions from the pythia and MadGraph simulations, where the widths of the bands indicate the uncertainties arising from the limited number of simulated events.}
\label{fig:dijet}
\end{center}
\end{figure}

ATLAS and CMS have measured the inclusive beauty cross section for {pp} collisions at $\sqrt{s}$ = 7 TeV by means of jets tagged by an algorithm using secondary vertex information with 2010 data~\cite{atlas_inc,cms_inc}. A displaced vertex is a good tag of the presence of a {b}-quark originated jet. The same measurement has been preformed using muons within the jets. As discriminating variable the transverse momentum relative to the jet axis ("p$_{\rm T}$ rel") has been used. In both experiments good agreement between the measurement and theory calculations is seen, as shown in Fig.~\ref{fig:inc}, except for some discrepancies at large p$_{\rm T}$ and $y$. The ATLAS result includes a measurement of the di-jet cross section, which also agrees with theory calculations, as seen in~\ref{fig:dijet} (left), except for low azimuthal angles ($\phi$) between the two b-jets. A similar discrepancy, seen in Fig.~\ref{fig:dijet} (right), was observed in a previous CMS result which studied the separation in the ($\eta$,$\phi$) plane ($\Delta$R) between the directions of the two b hadrons (BB), found using only information from the tracker.

\section{Conclusions}

There are many high quality results on heavy flavor production by both ATLAS and CMS. The observation of the new $\chi_{\rm b}$(3P) particle has been presented. The measurement of the exclusive $\Lambda_{\rm b}$ production cross section shows discrepancies with MC predictions. Open heavy flavor production is well described by theory calculations, although some discrepancies with the measurements can be seen at high p$_{\rm T}$ and , and at low di-jet separation angles.

\end{document}